# Beam characterization by means of emission spectroscopy in the ELISE test facility

M. Barbisan[1], F. Bonomo[2], U. Fantz[2] and D. Wünderlich[2]

[1]Consorzio RFX (CNR, ENEA, INFN, Univ. of Padova, Acciaierie Venete SpA), C.so Stati Uniti 4 – 35127, Padova (Italy)
[2]Max-Planck-Institut für Plasmaphysik (IPP), Boltzmannstr. 2, 85748 Garching, Germany

The ELISE test facility at IPP Garching hosts a RF H$^-$/D$^-$ ion source and an acceleration system. Its target is to demonstrate the performance foreseen for the ITER NBI system in terms of extracted current density (H/D), fraction of co-extracted electrons and pulse duration. The size of the ELISE extraction area is half that foreseen for the ITER NBI. This paper presents a detailed study of the ELISE beam divergence and uniformity. In particular, it was possible to describe the beam as the sum of two components at very different divergence: about 2° vs. 5°÷7°. As test cases, the beam properties have been measured as function of two source parameters. The first one is the current flowing through the grid facing the plasma, the Plasma Grid, in order to generate the magnetic filter field. The second one is the bias current flowing between the Plasma Grid and the source walls. Both the filter field and the bias current influence the fraction of co-extracted electrons, but also the properties of the plasma just in front of the extraction system and the beam properties.

The divergence and the uniformity of the beam have been measured by a Beam Emission Spectroscopy (BES) diagnostic; the detailed analysis of the raw spectra collected by BES led to describing the beam with two components of different divergence. This concept has been supported by the information given by thermal imaging of the diagnostic calorimeter. Further support to the proposed beam model has been found in the behavior of the currents flowing in the acceleration system and beamline components; these currents are given by the most divergent (charged) particles of the beam.

## 1. Introduction

The ELISE (Extraction from a Large Ion Source Experiment) test facility constitutes an important step towards the development of the ITER neutral beam injectors (NBIs). The facility hosts a RF negative ion source, coupled to an acceleration system with a maximum total acceleration voltage of 60 kV and half the extraction area (0.1 m$^2$) foreseen for the ITER NBIs [1][2]. The main targets of the research in ELISE are keeping the ratio of co-extracted electrons below 1 and reaching levels of extracted current density (285 A/m$^2$ for deuterium, 330 A/m$^2$ for hydrogen) and beam uniformity (>90 %) compatible with ITER requirements [3]. Because of limitations in the HV power supply, the extraction from the source can be performed for 10 s at intervals of about 150 s, while the plasma in the source can be sustained for up to 1 h [1][2][4].

The ion source in ELISE is composed by 4 RF plasma drivers, coupled to the acceleration system through a common expansion region. 2 ovens evaporate cesium in the source to increment the negative ion production by the surface process [5].

The acceleration system is composed by 3 ITER-like grids, namely the Plasma Grid (PG – facing the source), the Extraction Grid (EG) and the Grounded Grid (GG) [1]. To reduce the electron temperature in proximity of the PG, a horizontal magnetic filter field is generated by a current $I_{PG}$, made flowing vertically through the PG. As example, for a current of 2.5 kA a field of 2.4 mT is generated at 66 mm upstream the center of the PG [6]. The magnetic field contrasts the negative ion destruction by electron impact and reduces the amount of co-extracted electrons. In addition to $I_{PG}$ a current $I_{bias}$ is applied between the PG and the source walls together with the so called Bias Plate (BP) [1], which surrounds the PG apertures on the source side. The resulting voltage difference between BP and PG additionally reduces the amount of co-extracted electrons. At last, most of the co-extracted electrons are dumped on the EG surface, thanks to the magnetic field generated by the permanent magnets embedded in the EG itself [1].

The apertures of all the grids are 14 mm in diameter; the electrons and ions extracted from the apertures form beamlets. The apertures are grouped in 8 rectangular arrays of 5(h)×16(v), forming so-called beamlet groups; the beamlet groups are aligned in 2 horizontal rows, one above the other, with four groups each, and corresponding to the two sections of the grids [1]. The scheme of the PG is displayed in Figure 1a, with the 8 rectangular beamlet groups and the apertures. The projection of the four cylindrical drivers on the PG is given by red circles. The exit of the acceleration system is surrounded by a metal structure, formed by the so called Grounded Grid Holder Box and Green shield (GGGHB) [1], which acts as electrostatic shield for the beam.

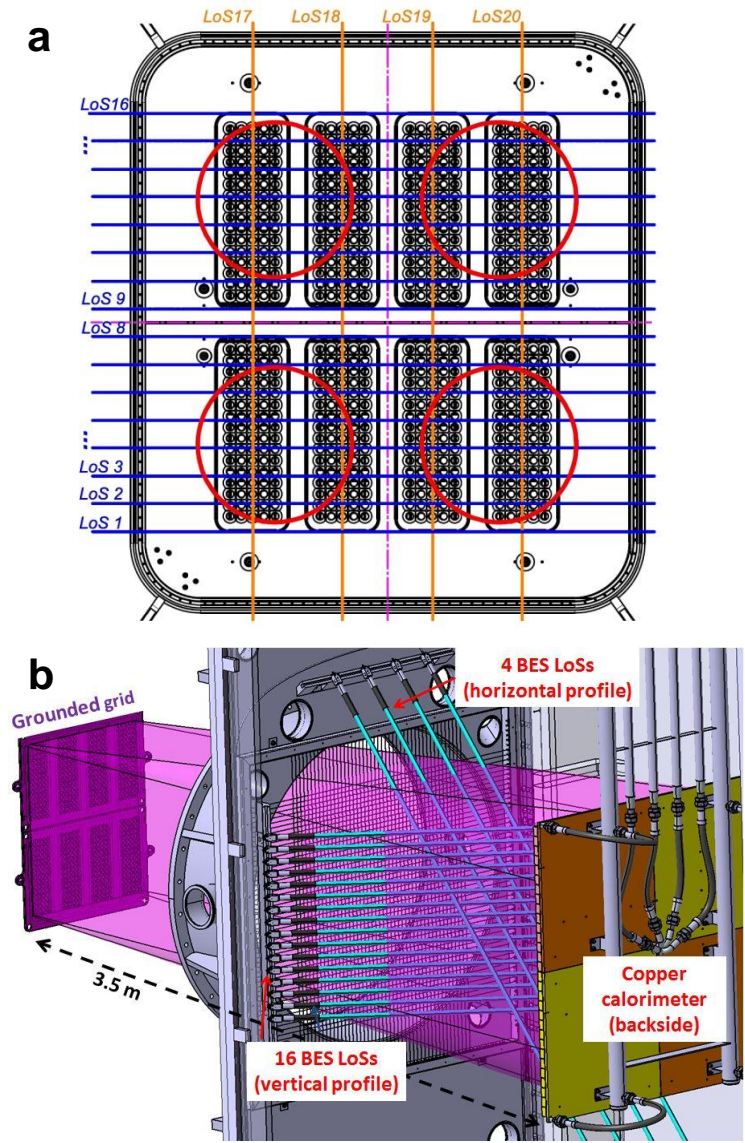

**Figure 1:** Picture a: layout of the apertures in the PG. The projection of the 4 plasma drivers is indicated with red circles. The blue and orange straight lines represent the projection of the horizontal and vertical Lines Of Sight dedicated to the Beam Emission Spectroscopy diagnostic. Picture b: 3D view of the BES horizontal and vertical optic heads, together with the calorimeter. The LoSs are indicated with light blue cylinders.

The presence of an horizontal magnetic filter field in front of the PG causes a vertical beam deflection due to the Lorentz force: in the current configuration of the filter field this deflection is directed downwards. In order to strengthen or weaken the field in front of the PG, external permanent magnets can be added to the source walls. This, however affects not only the intensity of the magnetic field but also the field topology close to the walls [6]. Also the beam deflection is affected by the presence of the external magnets.

In ELISE different diagnostics can be used to monitor the properties of the beam. The Beam Emission Spectroscopy diagnostic [7] can monitor the divergence and the uniformity of the beam, together with the fraction of negative ions which have been neutralized in the acceleration system (stripping losses). The beam divergence and uniformity can be derived also from the 2D map of the beam power deposited on the copper diagnostic calorimeter (shown in Figure 1b). This is divided in 4 quadrants, over which 900 copper blocks are installed in total. The blocks are connected to the quadrants but not between them, in order to reduce the thermal conduction along the plane of the calorimeter. The thermal footprint on the calorimeter can be studied by water calorimetry (one measurement system per quadrant), by means of thermocouples (embedded into 48 of the blocks) and IR thermography [7].

Due to the limit of the total high voltage power supply, the ITER NBI requirements in terms of beam divergence (< 0.4° for the beam core [8]) cannot be achieved in ELISE; the minimum divergence obtained in ELISE is about 1.5 °. In particular, due to the beamlet divergence combined with the large distance between the GG and the beam diagnostics, it is not possible to characterize the single beamlets. Thus, only averaged/global values for beam properties (both in divergence as in intensity or uniformity) can be derived. In particular, for the data presented in this paper, the source was not operated close to the minimum of the optimum optics, so that the values derived for divergence will be larger than 1.5°.

Besides the calorimeter and the BES diagnostic, it is possible to retrieve further information about the beam properties by means of the electrical measurements performed in the acceleration system. In particular, the measurements of currents flowing inside the grids and specific electrical shields give information on the most diffused and divergent part of the beam. The EG current essentially represents the electrons co-extracted from the source. The current on the GG is mainly due to the most divergent part of the beamlets or to secondary electrons, unable to pass through the GG apertures. At last, the current measured on the GGGHB is due to the impact of electrons produced by the neutralization of the beam, but also due to very angled beam ions.

The article describes the method currently used for the analysis of BES spectra (standard evaluation), together with an improved method which determines with more accuracy the angular distribution of the beam particles (par. 2). It is shown how, with the new method, it is possible to consider the beam produced in ELISE as the sum of two components, one much more divergent than the other. The hypothesis is further discussed and demonstrated in par. 3, by studying the spatial variations of the beam density and the beam divergence, as measured by the BES diagnostic. At last, the coherency of the hypothesis with the electric measurements in the acceleration system is presented in par. 4. The existence of a non-negligible and highly divergent component of the beam might have consequences for the development of future Neutral Beam Injectors, since too divergent beam particles would hit the beam line components causing severe heat loads on them.

## 2. BES data analysis methods

The Beam Emission Spectroscopy technique is based on the spectral analysis of the $D_\alpha/H_\alpha$ radiation produced in the interaction of the beam particles with the background gas. In ELISE, the light is collected in the beam drift region by 20 optic head. Since the magnetic filter field of the source is oriented in horizontal direction, the variations of the beam structure in the vertical direction are the most relevant. Thus, 16 of the 20 optic heads lay in a vertical array, equally distributed 5 cm one from the other and giving information on the vertical profile of the beam; the remaining ones lay in a horizontal array of 4 heads, equally distributed 16 cm each, giving information on the horizontal beam profile [7]. The Lines of Sight (LoSs) determined by the optic heads look at the beam in co-direction and intercept the beam at about 2.6 m downstream the GG. In Figure 1b it is possible to observe the 3D aiming of the LoSs (indicated with light blue cylinders) dedicated to BES; the projection of the horizontal and vertical LoSs on the grids is instead shown in Figure 1a, with blue and orange lines, respectively. A typical spectrum for a deuterium beam pulse collected along one of the LoSs is shown in Figure 2. Part of the $D_\alpha/H_\alpha$ radiation comes from excited atoms of the background gas hit by the fast beam particles and from reflected plasma light; these photons are observed at the nominal wavelength ($\lambda_0$=656.1032 nm for D, $\lambda_0$=656.2793 nm for H) because of the low speed of the atoms. Other $D_\alpha/H_\alpha$ photons come from excited beam particles; in this case the observed wavelength $\lambda'$ is noticeably different from $\lambda_0$, and can be calculated according to the Doppler shift formula [9]:

$$\lambda' = \lambda_0 \frac{1 - \beta \cos \alpha}{\sqrt{1 - \beta^2}} \quad (1)$$

where $\beta$ is the ratio between the speeds of the emitting particle and the light, and $\alpha$ is the angle between the beam axis and the mean direction of the photons entering the BES optic heads. In ELISE, the alignment of both horizontal and vertical LoSs is such that $\alpha$=130° [7]. As a consequence, in the spectra a Doppler shifted peak is present, shifted with respect to $\lambda_0$ towards higher wavelengths in accordance with the total energy gained by the fully accelerated particles (voltage difference between PG and GG). Additionally, a minor emission of Doppler shifted $D_\alpha/H_\alpha$ is found at intermediate wavelengths between $\lambda_0$ and the full energy Doppler peak. These photons come from stripping losses, i.e. negative ions which are at reduced energy because they were neutralized inside the acceleration system. The contribution of stripping losses is usually peaked at a wavelength corresponding to the extraction energy (voltage difference between PG and EG).

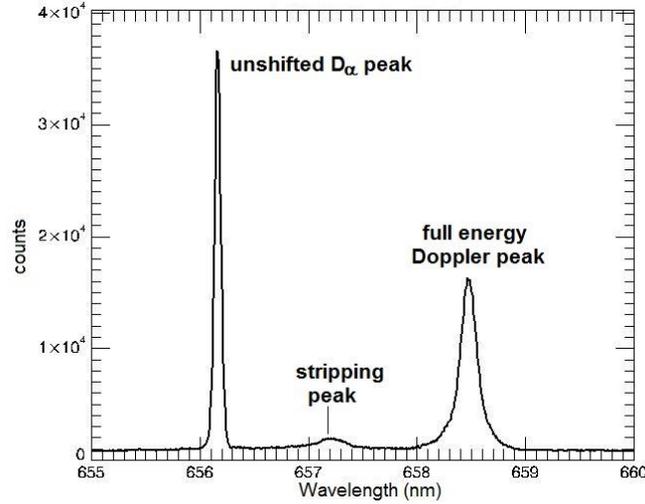

**Figure 2:** Typical spectrum acquired by the ELISE BES diagnostic (in deuterium).

In the analysis of BES spectra, it is possible to give a rough estimation of the beam profile uniformity by comparing the integral of the full energy Doppler peak from LoS to LoS. The beam divergence is calculated from the spectral width of the full energy Doppler peak: divergence can be considered as a fluctuation of the angle of observation $\alpha$ and then of $\lambda'$. Assuming that the transversal power density profile of beamlets is Gaussian, then also the Doppler peak is expected, in first approximation, to follow a Gaussian distribution. In the current BES analysis method at ELISE (standard evaluation method), the peak width is then measured by means of a Gaussian fit, applied only on the portion of the peak which is higher than the 30 % of the peak amplitude [10]. The reason of this choice is that the basis of the Doppler peak actually does not follow the Gaussian trend [11]. Once the sigma width $\sigma_D$ of the Doppler peak has been obtained, the e-folding divergence $\varepsilon$ of the beam is calculated according to the following formula:

$$\varepsilon = \sqrt{2} \cdot \sqrt{\frac{\sigma_D^2 - \sigma_S^2}{[(\lambda' - \lambda_0) \tan \alpha]^2} - \omega^2} \quad (2)$$

where $\sigma_S$ is the sigma width of the spectrometer instrumental function and $\omega$ represents a fluctuation in the collection angle due to the finite dimensions and the focusing of the optic heads. In the case of the BES diagnostic in ELISE, $\sigma_S$=17.9 pm and $\omega$=0.16°. Theoretically, eq. 2 must also include the broadening of the Doppler peak due to the ripple of the high voltage for the acceleration system, and then of the absolute values of particles' speed; in ELISE, however this effect is negligible. This type of analysis will be hereafter called "standard evaluation method".

From a spread check of the acquired BES spectra it resulted that the Doppler peak shows a broad component at its basis, which constitutes a non-negligible part of the peak itself. The broad component is in some cases so large that the threshold of 30 % in peak amplitude for the fit is not sufficient. Similar conditions were encountered in BES spectra collected from positive ion beams [12][13]; the adopted solution was to fit the full energy Doppler peak with a double Gaussian fitting function. Since in the spectra acquired in ELISE the Doppler peak often results to be slightly asymmetric, a slightly different fitting function has been proposed for the analysis. This function is the sum of a symmetric Gaussian n(λ) (narrow component) plus a bi-Gaussian curve b(λ) to fit the broad wings of the peak (broad component). n(λ) can be expressed as

$$n(\lambda) = A \cdot e^{-\frac{(\lambda-\lambda_{0n})^2}{2\sigma_{D1}^2}} \qquad (3)$$

where A is the peak level, $\lambda_{0n}$ is the center wavelength and $\sigma_{D1}$ is the sigma width. The bi-Gaussian curve, b(λ), is obtained joining together two halves Gaussian curves with different widths but same center and height:

$$b(\lambda) = \begin{cases} B \cdot e^{-\frac{(\lambda-\lambda_{0b})^2}{2\sigma_{D2}^2}} & if \quad \lambda \leq \lambda_{0b} \\ B \cdot e^{-\frac{(\lambda-\lambda_{0b})^2}{2\sigma_{D3}^2}} & if \quad \lambda > \lambda_{0b} \end{cases} \qquad (4)$$

Where B is the peak level of the bi-Gaussian, $\lambda_{0b}$ is its centroid and $\sigma_{D2}$ and $\sigma_{D3}$ are the sigma widths. The position of the center of the bi-Gaussian is a free parameter for the fit. The double width of the asymmetric Gaussian could account for the asymmetries of the Doppler peak. The fitting function does not account for the contribution of stripping losses to the low energy side of the full energy Doppler peak (i.e. at wavelengths lower than λ'); in the BES spectra collected in ELISE, however, this issue is normally negligible. Even if there are no theoretical indication of any asymmetry for the two beam components, the presence of a broad component (bi-Gaussian) centered at a higher wavelength with respect to the symmetric Gaussian of the narrow component suggests the presence of a different velocity angle between the particles contributing to the broad component and the line of observation with respect to the particles for the narrow component. This aspect is not yet clarified, and will be investigated furthermore in future. The new model fitted very well the experimental data, and allowed to separate the contributions of the 2 components to the total peak intensity. Figure 3 shows an example of the fitting curve: the total fitted function is indicated with at red dashed curve and is very well superimposed to the experimental data (in black). The bi-Gaussian curve representing the broad component of the peak is drawn with a blue dashed curve, while the symmetric Gaussian, i.e. the narrow component, is indicated with an orange dashed curve.

A divergence can be defined and evaluated for both the components of the fit. The divergence of the narrow component is estimated from the width $\sigma_{D1}$ of the narrow Gaussian, using eq. 2 (with $\sigma_{D1}$ in place of $\sigma_D$). The divergence of the broad component is instead estimated as the average of the values of divergence calculated separately from $\sigma_{D2}$ and $\sigma_{D3}$, using eq. 2 ($\sigma_{D2}$ and $\sigma_{D3}$ in place of $\sigma_D$).. The fraction of broad component in the beam is given by the ratio of the integral of the bi-Gaussian over the integral of the whole fitted function. The calculated value is however an underestimation over the real one, because the higher the divergence, the lower the fraction of beam volume intercepted by a LoS and then the lower the amount of photons collected by the diagnostic.

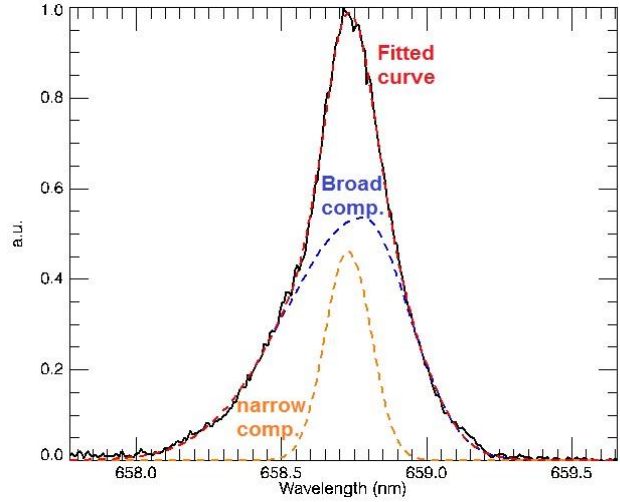

**Figure 3:** Example of fit of the full energy Doppler peak with the new function (Gaussian plus asymmetric bi-Gaussian). The black curve represents the measured spectrum (normalized to the peak intensity acquired during a hydrogen pulse, while the red dashed curve indicates the whole fitted function. The blue dashed curve indicates the bi-Gaussian curve, representing the broad component of the peak, while the orange dashed curve indicates the (symmetric) Gaussian, i.e. the narrow component of the peak.

### 3. Results in various operative scenarios

The validity of the new analysis method and of its physical assumption, i.e. the double angular distribution of beam particles, is firstly discussed by comparing the spatial variations of beam density and divergence. The measurements were obtained with both the standard evaluation method and with the new one. The results related to the 16 horizontal LoSs have been considered for this study. In this paper, investigations on the different wavelength between the narrow and the broader component are not reported.

2 series of pulses have been performed, varying the magnetic filter field (given by $I_{PG}$) or the bias current $I_{bias}$. $I_{bias}$ and $I_{PG}$ were varied in the intervals 30 A ÷ 60 A and 2.5 kA ÷ 3.9 kA, respectively; all the other parameters of source and acceleration system were left constant: 120 kW of input RF power, 0.6 Pa source pressure, 5.6 kV extraction voltage (potential difference between EG and PG) and 22 kV acceleration voltage (potential difference between GG and EG). All pulses were performed in deuterium, with permanent magnets attached on the source walls in such a way to strengthen the magnetic filter field (+0.4 mT at 66 mm upstream the center of the PG [7]).

BES results obtained with the standard evaluation method are shown in Figure 4. Plots a and c, respectively, show the vertical profiles of divergence and full energy Doppler peak integral, for the series of beam pulses in which $I_{bias}$ was varied. Plots b and d show the vertical profiles of divergence and full energy Doppler peak integral, for the scan in $I_{PG}$. In all the plots the "0" level in the axes of vertical position corresponds to the vertical center of the grids; as reference, the positions of the active optic heads are indicated with dashed lines.

The trends of the spatial profiles shown in Figure 4 are strictly related to the position of the beamlet groups. Taking into consideration the relation between the beam divergence and the beam intensity throughout the beam optics, divergence is higher in between the 2 rows of beamlet groups. In this region the beam intensity is lower, and is given by the overlapping of the beam intensity from the upper row segment with the intensity from the lower segment. The divergence increases strongly for the two topmost LoSs; a similar effect is not observed for the bottommost LoSs because of the slight downward orientation of the beam. The amplitude of the central peak in the divergence profiles increases with both the bias current and the PG current. Usually, to give a representation of the beam properties, only some LoS of the BES diagnostic are taken into account, namely those located in the projections of the drivers onto the grids (see Figure 1); these LoSs correspond approximately to the center of the beamlet groups, but taking into account also the vertical beam deflection.

The vertical profiles in Doppler peak intensity shown in Figure 4c and Figure 4d indicate that the beam is less intense in its bottom half; moreover, increasing $I_{bias}$ or $I_{PG}$ leads to a general further reduction of beam intensity, as well as for the total extracted current. This is due to the fact that not only the co-extracted electron current but also the extracted negative ion current is affected by increasing $I_{bias}$ and $I_{PG}$; for experimental operation usually a compromise between a large suppression of the electrons and a small decrease in the extracted negative ion current is used.

The trends of the Doppler peak intensity profiles are qualitatively coherent also with the observed thermal footprints of the copper calorimeter [7]. Differences in the pattern from BES profiles and the power deposited on the calorimeter can be attributed to the larger distance of the calorimeter from the GG: 3.5 m instead of around 2.6 m of the BES. The beamlet overlapping at the calorimeter is therefore always more pronounced than at BES location. Figure 5 shows the temperature pattern on the calorimeter surface, for the pulses of the scan in bias current corresponding to the lowest (picture a) and highest (picture c) currents. Pictures b and d refer instead to the pulses of the scan in $I_{PG}$ having the lowest and highest current, respectively. The temperature patterns, in agreement with the BES data, show that with higher bias current or PG current, the 2 rows of beamlet groups are less intense (especially the bottom row) and distinct. This pattern is found also in the temperature profiles from the thermocouples installed on the copper calorimeter as well as in the results of water calorimetry. The reason for different intensities among the different segments is not yet clear and is ongoing investigation. The plasma drift towards the top part of the source (less pronounced as observed in smaller negative ion sources like the prototype source used at the BATMAN test facility) could be an explanation.

The relation between the beam divergence profiles and the beam intensity profiles, given by BES or by the calorimeter, is that where the beam is more divergent (and then more diffused) the beam intensity is locally lower. This principle well explains the values of divergence and Doppler peak intensity at the vertical positions corresponding to of beamlet group position, as shown in Figure 4.

The beam core divergence is derived by the LoSs corresponding to the projections of the driver on the grid system, corresponding to about 2 ° for both the performed scans. Then, the divergence increases at positions which the Doppler peak intensity profiles indicate as the borders of beamlet groups, i.e. at beam edge and in the middle region between the upper and lower beamlet groups. The behavior of the beam divergence in these areas are explained as follows: those zones are not in the projection of the 2 rows of beamlet groups, so what BES LoSs observed are beam particles which have a wider angular distribution with respect to the "cores" of the beamlet groups, and which can then diffuse farther in transversal direction. The spatial profiles obtained by BES with the standard method then support the assumption of a double angular distribution for the beam particles.

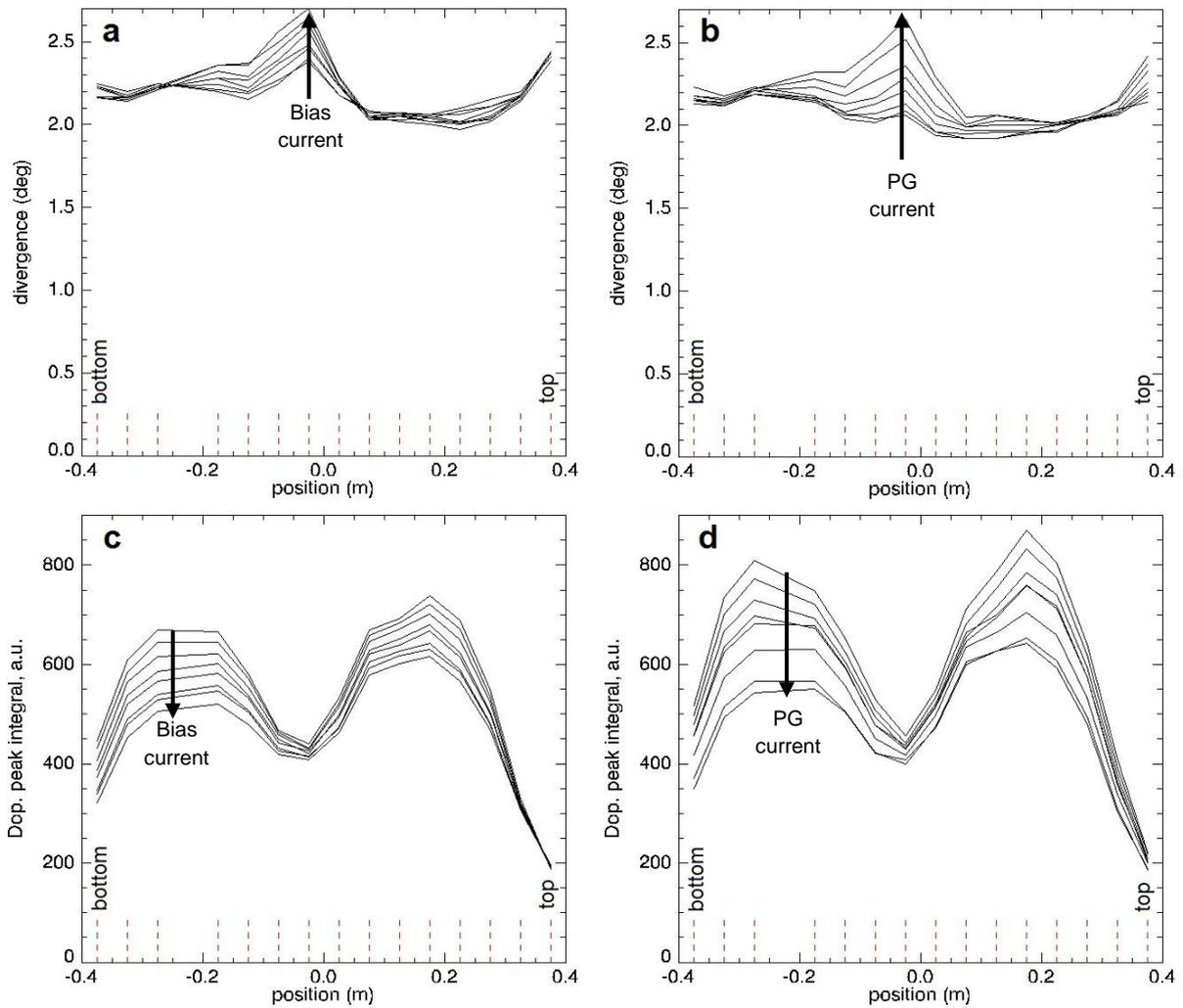

**Figure 4:** Beam divergence (plots a and b) and integral of the full energy Doppler peak (plots c and d), as function of the LoS vertical position, as from the standard evaluation method. The profiles shown in plots a and c belong to a scan in bias current (30 A→66 A), while those of plots b and d belong to a scan in PG current (2.5 kA →3.9 kA). Both the scans were performed in deuterium. In all the 4 plots the positions of the LoSs are indicated with dashed lines; the 4[th] LoS from the bottom is missing because it's out of order.

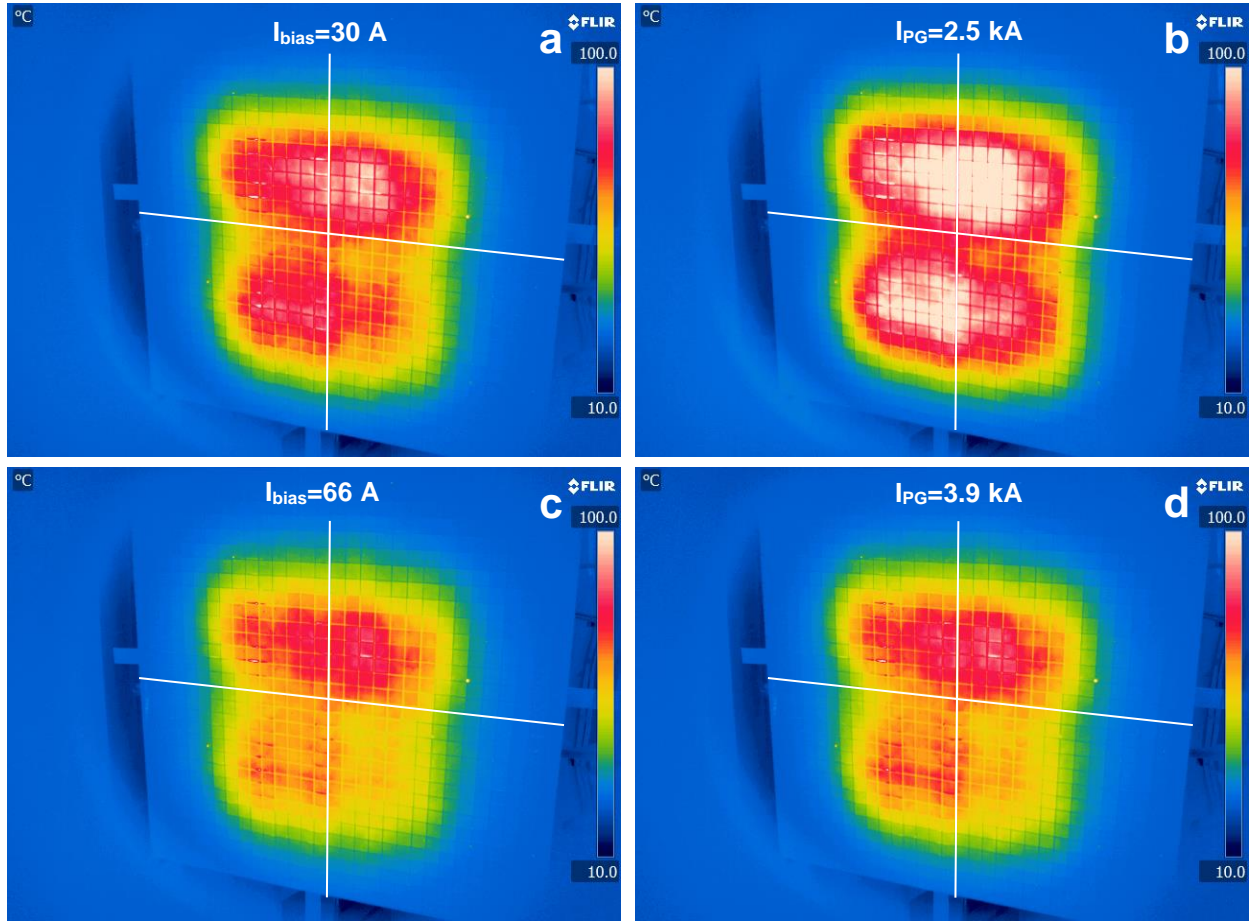

**Figure 5:** Thermal footprints of the diagnostic calorimeter, as measured by the IR camera. The colour scales are reported with the extreme values expressed in degrees Celsius. The horizontal and vertical axes, centered on the longitudinal axis of the source, are also reported. Pictures a and c belong to the pulses with respectively the lowest (30 A) and the highest (66 A) bias current in the scan described by plots a and c of Figure 4. Pictures b and d belong to the pulses with the lowest (2.5 kA) and the highest (3.9 kA) PG current in the scan described by plots b and d of Figure 4.

What emerged from the analysis of the BES spectra with the standard method is further clarified by using the new analysis method. Figure 6a shows the divergence attributed to the broad component of the beam, as a function of the vertical position of the LoSs; the profiles refer to the same scan in bias current described in Figure 4, plots a and c. The evaluated values of divergence are clearly higher than what obtained by the standard evaluation method (cp. Figure 4a); moreover, the divergence has substantially two levels, spread in space, whose positions can be associated to the top and bottom rows of beamlet groups. The divergence slightly increases with $I_{bias}$, by no more than 10 % within the selected data. By contrast, the values of divergence of the narrow component, shown in Figure 6c, are slightly lower with $I_{bias}$ (~-10 %) and the profile is almost flat.

The integral of the broad component, i.e. the integral of the bi-Gaussian, varies in the vertical direction as shown in Figure 6b, for the same pulses considered in Figure 5a. As it can be observed, in absolute values the integral of the broad component shows much lower spatial variations with respect to the profiles of Figure 4c; the contributions relative to the rows of beamlet groups can be barely distinguished, and sligthly grow with increasing bias current.

The integral of the broad component can be compared to the integral of the whole full energy Doppler peak. In Figure 6d the ratio between the integral of the bi-Gaussian and the integral of the whole fitted curve is reported as function of the LoS vertical position, for the same pulses of plots a and b. What is shown is that the fraction of the broad component increases with the bias current, and is peaked in between the rows of beamlet groups. In substance, the integral of the broad component (the bi-Gaussian) is uniform over the beam region, because of the high level of divergence of the broad component itself; vice versa, the intensity of the narrow component (the symmetric Gaussian) is much more localized in correspondence of the rows of beamlet groups. It must be noticed also that the broad component can account for up to the 70 % of the entire Doppler peak intensity in the beam central region. This explains the typical shape of the divergence spatial profiles obtained by the standard evaluation method and shown in Figure 4: even fitting the Doppler peak from the 30 % of its peak value, the influence of the underlying broad component cannot be always avoided: where it is dominant (i.e. in between the rows of beamlet groups), the calculated divergence is systematically increased. It also explains why the divergence increases in the top part of the beam for all the profiles shown in this paper: being more divergent, the broad component of the beam has higher spatial dimensions, therefore at the beam edges it constitutes the major fraction of the beam itself (as shown in Figure 6d). Same consideration should account for the bottom part of the beam, which is however not visible to the lowest BES LoSs.

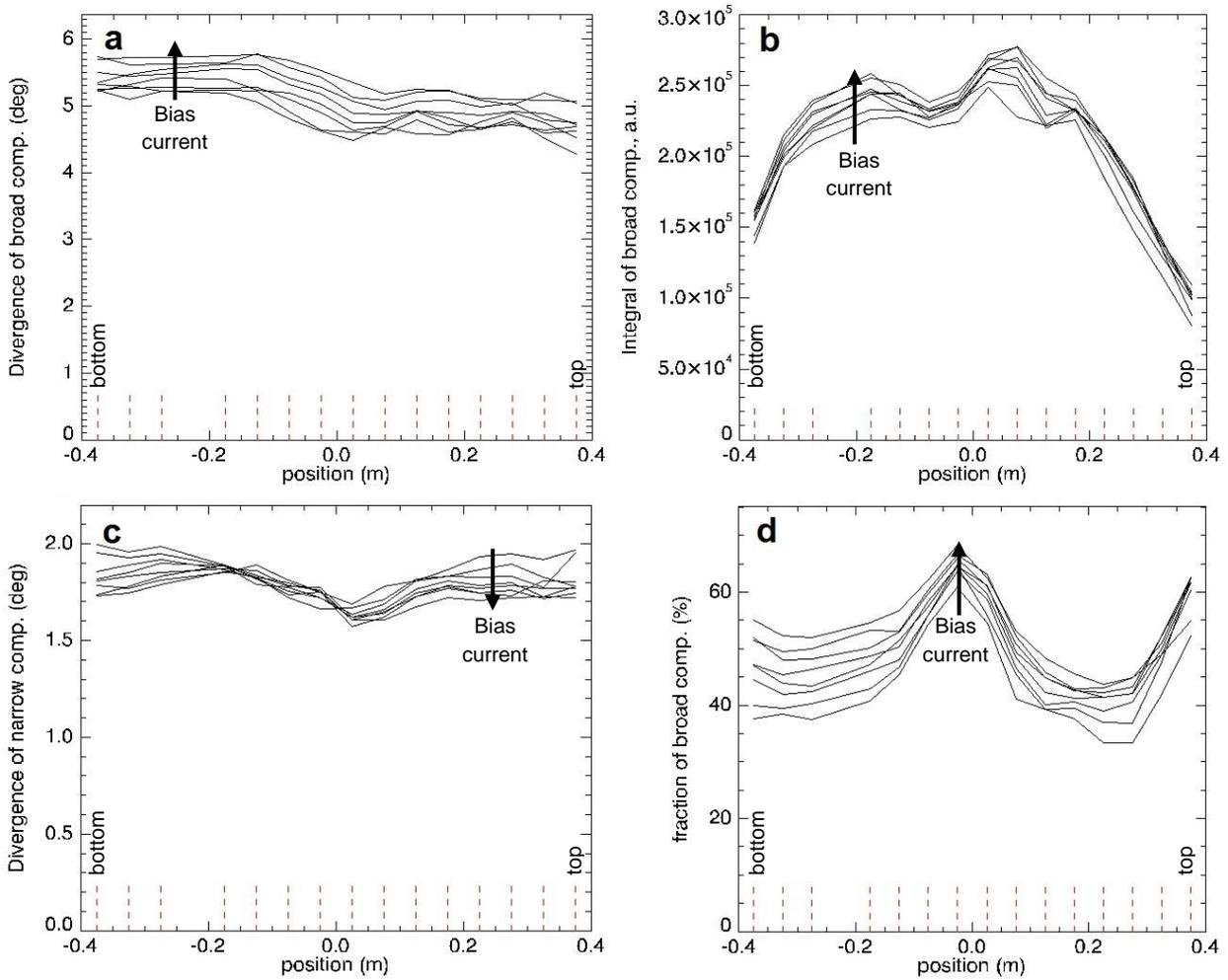

**Figure 6:** Various properties of the broad component of the full energy Doppler peak, estimated with the new analysis method and plotted as function of the LoS vertical position. The average of the divergences calculated from the widths of the bi-Gaussian is shown in plot a. Plots b and d respectively show the integral of the bi-Gaussian and its ratio with the integral of the whole fitted function. Plot c shows the divergence calculated from the narrow symmetric Gaussian. The profiles of the 4 plots belong to the same scan in bias current considered in Figure 4. As reference, the positions of the LoSs are indicated with dashed lines; one LoS (the 4th from the bottom) is missing because it's out of order.

In addition, the narrow component divergence (Figure 6c) is almost flat at around 1.7 deg, a value 10% smaller than the one estimated in the standard evaluation, suggesting that the vertical divergence profile pattern (higher divergences in correspondence of the beam edges or in between the 2 beamletgroup rows) is essentially defined by the broad component.

In conclusion, the spatial variations of the data given by the new analysis method support the assumption that the beam has a broad component. It is also clear that the new analysis method gives more accurate results with respect to the standard method, especially in the regions of the beam which are far from the core of the beamlet groups.

4. Comparison with electrical measurements

The existence of a beam component with a large divergence as described above might represent a problem, not for ELISE itself but in perspective for the future neutral beam injectors [14]. The power deposited on the beam line components will be strictly related to the beam divergence: the higher the divergence, the higher the heat load. The existence of a broad component of the beam has been further verified in ELISE by studying the currents flowing on the beam line components which may be intercepted by strongly divergent beams. The study was carried out on hydrogen pulses performed with different magnetic filter field amplitude (varied by acting on $I_{PG}$); as for the cases studied in the previous chapter, the scan was performed with permanent magnets installed on the sides of the source body and strengthening the field produced by $I_{PG}$ on the source side. Except for $I_{PG}$ (higher values are needed for deuterium operation with respect to hydrogen pulses), all the other source parameters were identical to the ones used in the previous section. Additionally, the extraction and acceleration voltages have been modified: $U_{ex} = 7.6$ kV, $U_{acc} = 25$ kV. The extracted current is slightly higher for hydrogen pulses than for the deuterium ones and can be due also to a different caesium conditioning status of the source during the experiments for both isotopes.

Figure 7a shows the total current $I_{EG}$ collected by the Extraction Grid, as a function of $I_{PG}$; $I_{EG}$ is mainly due to co-extracted electrons and therefore, as expected, it falls with increasing filter field amplitude. For the selected beam pulses, the electron current density is about 30 % of the extracted ion current density (~110 A/m$^2$), therefore it contributes negligibly to the beam optics. Downstream the EG, the extracted ion current, $I_{ion}$, can be collected by the surfaces of several components:
- the Grounded Grid;
- the Grounded Grid Grid Holder Box and the Green Shield (GGGHB);
- the calorimeter and the vacuum vessel, which are electrically connected.

The currents on the GG and the GGGHB, $I_{GG}$ and $I_{GGGHB}$, can be measured separately. The fraction of the extracted beam current collected by the GG, i.e. $I_{GG}/I_{ion}$, is shown for the considered cases in Figure 7a too, as function of $I_{PG}$. Contrarily to the behavior observed for $I_{EG}$, a higher fraction of the beam impinges on the GG when the magnetic filter field is increased; this could be explained by a broadening of the beamlets inside the acceleration system, which is coherent with what observed by BES. The same explanation can be given to the behavior of the current flowing in the GGGHB. Figure 7a shows also the fraction of the beam current downstream the GG and collected on the GGGHB (i.e. $I_{GGGHB}/(I_{ion}-I_{GG})$), plotted as function of the PG current. It's evident that increasing the amplitude of the magnetic filter field raises the fraction of beam charges which are sufficiently angled to hit the GGGHB.

The electric measurements can be correlated to the results of the new analysis method for the BES spectra. It is found that, for the considered pulses in hydrogen, the hypothetical broad component of the beam has a divergence of about 7.2°, increasing with $I_{PG}$ by no more than 10 % for the considered scan. However, the fraction of broad component over the whole beam increases together with the fraction of the beam current collected by the GGGHB. This is shown in Figure 7b, where the fraction of the broad component is plotted against $I_{GGGHB}/(I_{ion}-I_{GG})$. In detail, the values in Figure 7b indicated with blue diamonds are the average of the values obtained from the LoSs which intercepted the projection of the 2 rows of beamlet groups. For completeness, Figure 7b shows also the fraction of broad component averaged from the spectra of the 4 vertical LoSs (red squares). According to the shown values, the relative increase of the currents on GG and GGGHB could be explained with an increase of the fraction of the broad component in the beam, rather than with a general increase of the beam divergence. The gap between the values given by horizontal and vertical LoSs might be due to the fact that the top and bottom rows of beamlet groups, both observed by each vertical

LoS, often have slightly different directions in the vertical plane, with consequences for the shape of the full energy Doppler peak.

The relation between the current on the GGGHB and the broad component of the beam is further witnessed in Figure 8. The picture shows the Grounded Grid Grid Holder Box as looking from the source, during the disassembly of source and acceleration system in February-March 2015 for special maintenance; the Green Shield is barely visible, downstream the GG Grid Holder Box. During the operation of the source some copper had been sputtered from the grids and then deposited on the GGGHB surfaces, except for some areas whose position corresponds to that of the 2 rows of beamlet groups. It has been calculated that a particle of the outermost beamlets should travel with a horizontal angle of 7.7° to hit the downstream border of the Grounded Grid Grid Holder Box. To hit the upstream side of the clean areas shown in Figure 8, instead, ions should have travelled with an horizontal angle of about 10°÷11°. The e-folding divergence measured for the broad component of the Doppler peak is for the data here reported typically around 5° for deuterium and 7° for hydrogen. The isotope effect could play a role in the different divergences observed, but also other parameters such as the different grid voltages (i.e. beam optics) and the status of the caesium conditioning. No definite conclusions can be derived and further investigations are needed. In the horizontal direction these values of divergence are therefore sufficient to cause some beam particles to hit the Grounded Grid Grid Holder Box. The values of divergence obtained by the standard evaluation method are roughly half of those attributed to the broad component with the new method, and could not explain what observed on the Grounded Grid Grid Holder Box.

The data then support the hypothesis that a broad component of the beam really exists, and is partially intercepted at the beam borders by the GG Grid Holder Box.

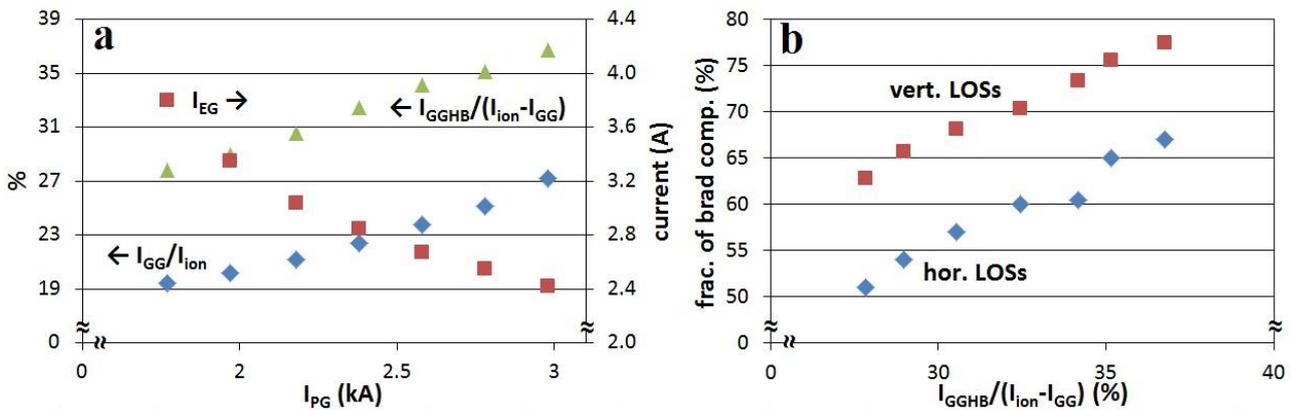

**Figure 7:** Study of the currents flowing on the EG ($I_{EG}$), the GG ($I_{GG}$), the Grounded Grid Holder Box and the Green Shield ($I_{GGGHB}$), in correlation with the BES data, for hydrogen pulses performed by varying $I_{PG}$. Plot a: $I_{EG}$, fraction of the extracted current which has been collected by the GG ($I_{GG}/I_{ion}$), fraction of the current flowing downstream the GG (i.e. $I_{ion}-I_{GG}$) which has been collected by Grounded Grid Grid Holder Box and Green shield, against $I_{PG}$. Plot b: fraction of the full energy Doppler peak belonging to the broad component. The values indicated with blue diamonds are calculated from the horizontal LoSs which intercepted the rows of beamlet groups; those indicated with red squares are instead calculated from the 4 vertical LoS. All the values are plotted against the fraction of the accelerated current collected by the GGGHB structure.

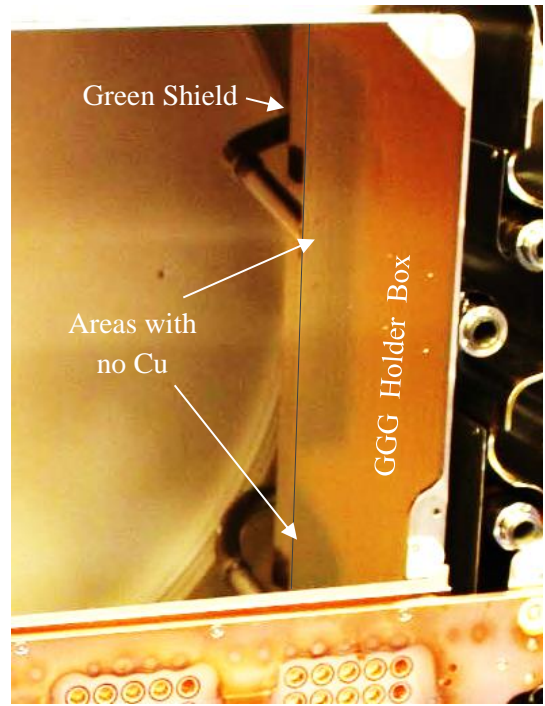

**Figure 8:** Picture taken during the opening of the ELISE source for special maintenance, in the period February-March 2015; the viewer is looking in the beam direction from the source. The vertical surface is the Grounded Grid Holder Box, just beyond it the Green Shield can be seen; the copper grid in the bottom is a section of the EG. As pointed out in the picture, the stainless steel Grounded Grid Grid Holder Box has been covered by copper (of the grids) during the ELISE operation, except for some areas corresponding to the rows of beamlet groups.

## 5. Conclusions

From the observation of the full energy Doppler peak in BES spectra a new analysis method has been proposed to analyze the spectra and characterize spatial variations of beam divergence and beam intensity. The new method is based on the assumption that the beam particles can be grouped in 2 components, one much more divergent than the other.

The new analysis method resulted to be more accurate than the standard evaluation method, in particular for the lines of sight which are peripheral with respect to the beamlet groups. In these cases, indeed, the tails at the base of the Doppler peak are sufficiently large to affect the results of the single Gaussian fit of the standard method.

Both the information given by the BES diagnostic and by the electric measurement on the acceleration system has confirmed that the beam has a broad component. The experimental evidences which support this assumption are in the following:

- The integral of the broad component of the Doppler peak shows very smooth variations in the vertical direction, qualitatively consistent with the high values of divergence attributed to the broad component.
- The fraction of the broad component composing the Doppler peak is sensible to variations of the bias current or of the magnetic filter field. The shape of the base of the Doppler peak is therefore not a product of some photons collection phenomenon, but represents a real feature of the beam particles' distribution.
- The currents flowing in the GG and the GGGHB, considered in proportion to the beam current fully accelerated and impinging on the calorimeter, increase when the fraction of the broad component on the Doppler peak is higher.
- The divergence attributed to the broad component is compatible with the high angular deviations necessary for the ions to hit the GGGHB structure. By contrast, the values of divergence given by the standard evaluation method are not compatible with the phenomena observed on the surface of the Grounded Grid Holder Box.

The existence of a consistent (> 40%) fraction of the beam which is highly divergent ($\varepsilon > 5°$) represents a risk for the safety of future neutral beam injectors based on negative ions. Possible measures for reducing the broad component (e.g. improving the beam optics) will be investigated. Further information about the broad component of the beam, more specifically about its dependency on the magnetic filter field, have been reported in [15].

The experimentation on the ELISE test facility will allow to understand the causes of the broad component formation. It has to be pointed out that ELISE was not operated at the global optimum of the beam optics during these experiments. The broad component might then be the result of the not optimized focusing of the negative ions inside the acceleration system. Preliminar tests, performed by varying the extraction voltage, indicate that the fraction of broad component of the beam is minimized (below 10%) when the beam perveance is such to minimize the overall beam divergence. Further detailed investigations on beam properties in an operational scenario closer to the global optimum perveance will be performed in order to draw a final conclusion for the ITER NBI systems.

**Acknowledgements**

The work was supported by a contract from Fusion for Energy (F4E-2009-0PE-32-01), represented by Antonio Masiello. The opinions expressed herein are those of the authors only and do not represent the Fusion for Energy's official position.